\documentclass[11pt,twoside,A4]{article}
\usepackage{times,fancyhdr}
\usepackage[dvips]{graphicx}
\usepackage{latexsym}
\usepackage[affil-it]{authblk}
\usepackage{mathptm}
\usepackage{amsmath}
\usepackage{amssymb}
\usepackage{color}
\usepackage{setspace}
\usepackage{hyperref}
\usepackage{floatrow}

\usepackage[top=4cm, bottom=4cm, left=4cm, right=4cm]{geometry}

\def\beq{\begin{equation}}
\def\eeq{\end{equation}}
\def\beqn{\begin{eqnarray}}
\def\eeqn{\end{eqnarray}}

\newcommand{\be}{\begin{equation}}
\newcommand{\ee}{\end{equation}}
\newcommand{\bea}{\begin{eqnarray}}
\newcommand{\eea}{\end{eqnarray}}

\begin{document}

\title{The Redshift-Dependence of Radial Acceleration: \\Modified Gravity versus
Particle Dark Matter}
\author{Sabine Hossenfelder, Tobias Mistele}
\affil{\small Frankfurt Institute for Advanced Studies\\
Ruth-Moufang-Str. 1,
D-60438 Frankfurt am Main, Germany
}
\date{}
\maketitle
\vspace*{-0.3cm}

\begin{abstract}
 
Modified Newtonian Dynamics has one free parameter and requires an
interpolation function to recover the normal Newtonian limit. We here show 
that this interpolation function is unnecessary in a recently proposed covariant completion
of Erik Verlinde's emergent gravity, and that Verlinde's approach moreover fixes the 
function's one free
parameter. The so-derived correlation between the observed 
acceleration (inferred from rotation curves) and the gravitational
acceleration due to merely the baryonic matter fits well with data.
We then argue that the redshift-dependence of galactic rotation curves could offer a way to tell apart
different versions of modified gravity from particle dark matter.
\bigskip

\end{abstract}

\section{Introduction}
 
It has been known for several decades that Modified Newtonian Dynamics ({\sc MOND}) \cite{Milgrom}
explains some observed phenomena that have proved difficult to reproduce with
particle dark matter. The most notable of these phenomena is the stunningly tight correlation between
the gravitational pull which we observe acting on stars in galaxies -- for example
through rotation curves -- and the gravitational pull that is caused by the `normal' 
baryonic matter only \cite{McGaugh}. While particle dark matter does a better job with the
temperature fluctuations of the cosmic microwave background and its flexibility
is of advantage to describe galaxy clusters, the successes of {\sc MOND} on
galactic scales are
remarkable regardless \cite{Famaey:2011kh}.

{\sc MOND} is a non-relativistic theory to which several relativistic completions are
known \cite{Rel1,Milgrom:2009ee,Bekenstein:2004ne,Zlosnik:2006zu,Milgrom:2009gv,Deffayet:2011sk,Blanchet:2015bia,Hossenfelder:2017eoh}. We 
here focus on a new member of this class, Covariant Emergent Gravity (CEG), which was recently proposed in \cite{Hossenfelder:2017eoh}. We will show that {\sc CEG}, unlike {\sc MOND},
does not need an interpolation function but instead
predicts a particular interpolation function. Moreover, if one follows the
argument proposed in \cite{Verlinde:2016toy}, this interpolation function has
no free parameters. We demonstrate that the so-obtained equation fits the
data well.


\section{Modified Newtonian Gravity}

The defining equation of {\sc MOND} is
\beqn
\vec \nabla \cdot \left( \mu (| \vec \nabla \phi_{\rm MOND}/a_0 |) \vec \nabla \phi_{\rm MOND} \right) = 4 \pi G \rho ~, \label{MOND}
\eeqn
 where $G$ is Newton's constant, $\rho$ is the energy density of baryonic matter and
$\phi_{\rm MOND}$ is the modified Newtonian potential. The function $\mu$ is the interpolation
 function and $a_0$ quantifies an acceleration that is the theory's one free parameter. From
 comparison with data one finds that the following relation is approximately correct \cite{Milgrom:2017hpv}:
\beqn
 2 \pi a_0 \approx H_0 \approx \sqrt{\Lambda/3}~,
 \eeqn
where $H_0$ is the Hubble constant and $\Lambda$ the cosmological constant. The numerical
 value is $a_0 \approx 10^{-10}$ m/s${}^{2}$.
 
The interpolation function is necessary to switch off the {\sc MOND} effects and get back normal
 General Relativity in deep gravitational potentials.  It
 is often taken to be of the form
 \beqn
 \mu (x) = \frac{x}{1+x}~,
 \eeqn
which does a good job to fit data on galactic sizes.

From $\mu$ one can calculate the relation between the total acceleration, $g_{\rm tot}$, and the acceleration created by the
baryonic (``normal'') mass only, $ g_{\rm B}$. This gives 
\beqn
\mu (  g_{\rm tot} /a_0) \vec g_{\rm tot} = \vec g_{\rm B}~,
\eeqn
where $ g_{\rm tot} = | \vec g_{\rm tot} |$.
In the limit where $1 \ll g_{\rm tot}/a_0$, ie 
accelerations are large compared to $a_0$, the 
interpolation function goes to 1, so we recover the normal Newtonian limit. On the other hand, in the limit where $ g_{\rm tot}/a_0 \ll 1$, ie
accelerations are small compared to $a_0$, we are in the {\sc MOND} regime where $g_{\rm tot}^2 = a_0 g_{\rm B}$. 

In the case of spherical symmetry, the solution to Eq.\ (\ref{MOND}) in the {\sc MOND}-regime is $\phi_{\rm MOND} \sim \ln (r)$,
as opposed to $\phi_{\rm MOND} \sim 1/r$ in the Newtonian limit. This means that in the {\sc MOND}-regime the force
acting on test-particles orbiting a large mass $M$ (eg the galactic center) is proportional to $1/r$, resulting in flat
rotation curves. It also follows that the total mass $M \sim v^4$, where $v$ is the limiting
velocity of the rotation curves. This is the observationally well-established
Tully-Fisher relation \cite{TF}.

\section{Covariant Emergent Gravity}

The defining Lagrangian of {\sc CEG} is that of matter coupled to gravity and an additional vector field $u_\mu$.
In the non-relativistic limit it gives rise to the equation \cite{Hossenfelder:2017eoh}
\beqn
\vec \nabla \cdot \left( \left|\vec \nabla \phi \right|  \vec \nabla \phi  \right) &=& \frac{2 \pi G}{3 L}  \rho ~, \label{CEG}
\eeqn
where $L$ is a constant of dimension length (more about which later), and $\phi$ is
proportional to the absolute value of the vector-field $u_\mu$. 

At first sight, Eq.\ (\ref{CEG}) looks pretty much the same as Eq.\ (\ref{MOND}) except for the
different constants. But they are not the same because the scalar $\phi$ in Eq.\ (\ref{CEG}) is
not the gravitational potential as in (\ref{MOND}). Instead, this scalar causes an additional force acting on baryons by direct interaction.
In {\sc CEG} the normal gravitational potential $\phi_{\rm N}$ is instead determined, as usual, by
\beqn
\Delta \phi_{\rm N} &=& 4 \pi G \rho ~. \label{CEG2}
\eeqn

In {\sc CEG} now the total acceleration, $g_{\rm tot}$, which acts on baryons comes from the gradient of
$\phi + \phi_{\rm N}$, not from the gradient of $\phi$ alone, as in {\sc MOND}. For a test-particle in a spherically symmetric
field, (such as a star orbiting the galactic center), the total acceleration is 
\beqn
g_{\rm tot} = g_{\rm B} + g_{\rm \phi}\quad \mbox{where} \quad g_{\rm B} = \partial_r \phi_{\rm N}~, ~g_{\rm \phi} = \partial_r \phi ~.\label{gceg}
\eeqn

In the case of spherical symmetry, it is straight-forward to use Eq.\ (\ref{gceg}) to calculate the relation between the acceleration
expected only from the baryonic mass, $g_{\rm B}$ and the total
acceleration $g_{\rm tot}$. In spherical coordinates, Eq.\ (\ref {CEG}) reads
\beqn
\frac{1}{r^2} \partial_r \left(r^2 \left(\partial_r \phi \right)^2 \right) &=& \frac{2 \pi G}{3 L}  \rho ~. \label{CEGspherical}
\eeqn
Multiplying this equation with $r^2$ and integrating it once yields
\beqn
  r^2 \left(\partial_r \phi \right)^2  = \frac{G}{6 L}  M(r) \quad \mbox{with} \quad M(r) = 4 \pi \int_0^r dr' ~ r'^2 \rho(r')~.
\eeqn
If we now divide this equation by  $r^2$, then the right side becomes just the normal gravitational acceleration
of the baryonic mass, $g_{\rm B}$, while the left side is the square of the acceleration that comes from the interaction
with the new field $g_{\rm \phi}$. Taking the square root, we get
\beqn
 g_{\rm \phi} =  \sqrt{ \frac{g_{\rm B}}{6 L}}~.
\eeqn
This can now be inserted into (\ref{gceg}) to give the `radial acceleration relation':
\beqn
  g_{\rm tot} = g_{\rm B}\left(1 + \sqrt{ \frac{\tilde a_0}{g_{\rm B} }} \right)~, \label{gtotb}
\eeqn
where $\tilde a_0 := 1/(6L)$. Finally, by solving Eq.\ (\ref{gtotb}) for $g_{\rm B}/\tilde a_0$, we see that in the non-relativistic limit {\sc CEG}  corresponds to {\sc MOND} with the interpolation function
\beqn
\tilde \mu (x) = (1 + 2x - \sqrt{1+4x}) / (2x) ~.
\eeqn

Relation (\ref{gtotb}) was previously mentioned in \cite{Milgrom:2016huh}. The derivation we
have presented here differs from the argument in \cite{Milgrom:2016huh} in two important points. First,
our derivation is valid for general, spherically symmetric mass distributions and not merely for a point
mass, as in \cite{Milgrom:2016huh}. Second, our result follows directly from a Lagrangian
formulation and not from ad-hoc equations. 

Let us then say something about the free constant $L$ which enters $\tilde a_0$. In \cite{Verlinde:2016toy}, Verlinde
fixes this constant by the following argument, hereafter referred to as `Verlinde-matching.' The additional force
acting on baryonic matter is caused by the change in entanglement entropy induced by the presence
of the matter. This change comes about because inserting a baryonic mass into an asymptotic de-Sitter space  
slightly shifts the de-Sitter horizon, thereby changing the volume inside the horizon. Verlinde then
requires that the horizon-shift induced by the presence of baryonic matter is identical to the shift quantified by the new
field, which leads to $ 1/L = \sqrt{\Lambda / 3} $ in a universe with $ \Omega_\Lambda = 1 $ and $ \Omega_m = 0 $, and $ 1/L \approx 1.05 \times \sqrt{\Lambda / 3} $ in a universe with $ \Omega_\Lambda = 0.7 $ and $ \Omega_m = 0.3 $.

While this argument lacks rigor, the consequence is that in the
non-relativistic limit, {\sc CEG} with Verlinde-matching 
has {\it no} free parameters. 

\section{Comparison with Observation}

Since a model without free parameters is every phenomenologist's nightmare, 
we now perform a sanity check and compare the radial acceleration relation (\ref{gtotb}) with observation. For this we use the
 data-set compiled in \cite{McGaugh} which collects 2693 measurements from rotation curves of 153 galaxies. 


\begin{figure}[h]
\hspace*{-1.5cm}\vspace{0cm}
\floatbox[{\capbeside\thisfloatsetup{capbesideposition={right,top},capbesidewidth=6cm}}]{figure}[8cm]
{\caption{Observed, total acceleration ($g_{\rm tot}$) versus acceleration due to baryonic mass only ($g_{\rm B}$). Blue
squares are data from \cite{McGaugh}. Red, solid curve: {\sc CEG} with Verlinde-matching. Pink shading: 1 $\sigma$ uncertainty. Dashed, black line: Newtonian gravity
without dark matter.  \label{fig1} }}
{\vspace*{-1.2cm} \includegraphics[width=8cm]{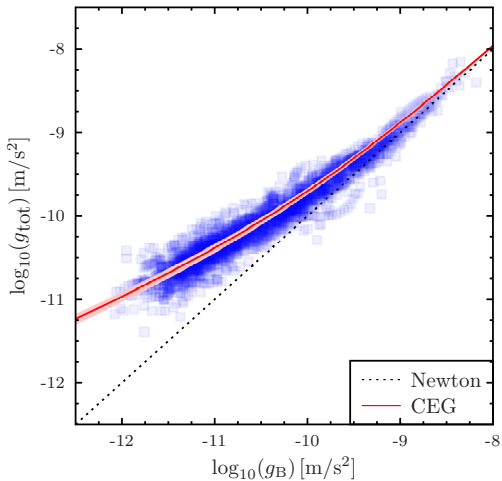} \vspace*{-0.5cm}}
\end{figure}
\vspace*{-0.5cm}


 For Figure \ref{fig1}
we have fixed $L$ using the Verlinde-matching as explained above. 
This gives the value $\tilde a_0 = (0.96 \pm 0.01) 10^{-10}$ m/s$^2$ 
with the dominant error coming from the uncertainty in the Hubble rate \cite{Planck:2015}. We note in the passing that the equation derived here 
from {\sc CEG} does not suffer from the problem with Verlinde's model pointed out in \cite{Lelli:2017sul}.

A $\chi^2$ fit gives the best-fit value $\tilde a_0 =  (0.77 \pm 0.01) \times  10^{-10}$ m/s$^2$, where
the uncertainty is that of the statistical fit. We do not plot the curve with the best-fit value because by eye it 
cannot be distinguished from the curve with the Verlinde-matching. The statistical uncertainty on $\tilde a_0$ 
is small due to the large number of data points, but the data bring in a measurement error
of $\sim 0.24  \times 10^{-10}$ m/s$^2$ from the normalization of the stellar mass-to-light ratio \cite{McGaugh}, which is the error depicted in Figure \ref{fig1} . We
conclude that {\sc CEG} with Verlinde-matching is consistent with data, at least so far.

We interpret {\sc CEG} as a limit in which the additional field is in a superfluid
phase. As laid out in \cite{Berezhiani:2015bqa}, this limit no longer
applies on the scale of solar systems (because the gradient of the field is too large) and
also not on the scale of galaxy clusters (because the average potential isn't deep enough).
Hence, one should not read too much into the fit at the lowest and highest accelerations.


\begin{figure}[th]
\vspace*{-1cm}

\floatbox{figure}[\textwidth]{\caption{Redshift-dependence of radial acceleration relation for {\sc CEG} and {\sc MOND}.  \label{fig2}}}{ \hspace*{-.3cm}\includegraphics[width=8cm]{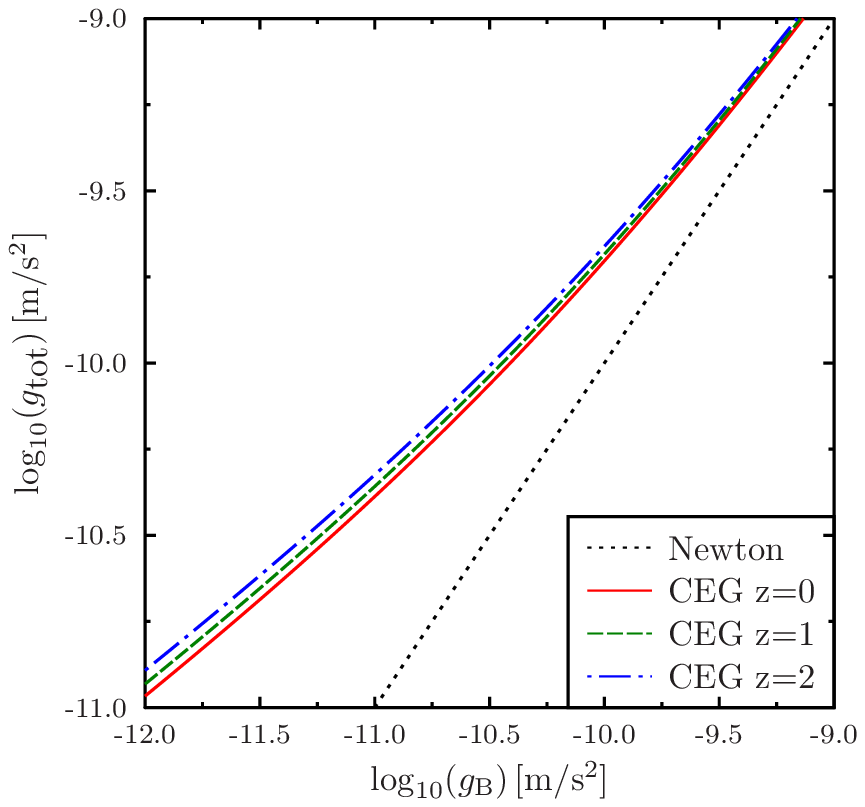}\hspace*{-1cm}\includegraphics[width=8cm]{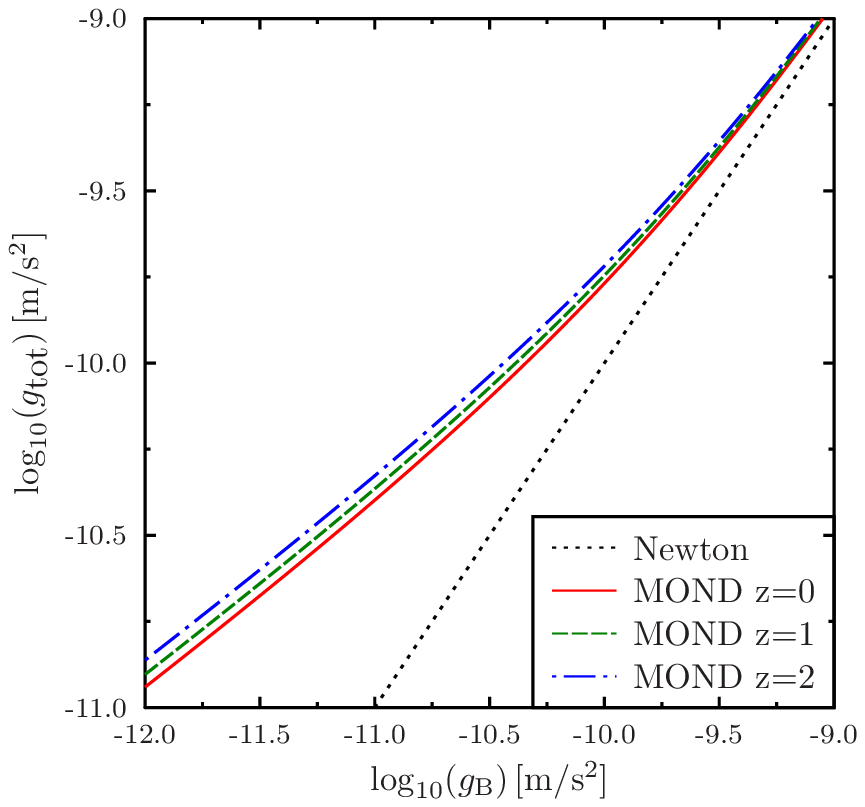} }

\end{figure}


Interestingly, however, as laid out earlier, Verlinde-matching relates $\tilde a_0$ with the size of the cosmological
horizon which is a redshift-dependent quantity. We expect such a redshift-dependence in
any approach that ties the acceleration scale to the de-Sitter temperature, 
as for example through the argument in \cite{Smolin:2017kkb}. It was even argued in \cite{Milgrom:2017hpv} 
that {\sc MOND} itself should have a redshift dependence.

\begin{figure}[th]
\hspace*{-1.5cm}\vspace{0cm}
\floatbox[{\capbeside\thisfloatsetup{capbesideposition={right,center},capbesidewidth=6cm}}]{figure}[8cm]
{\caption{Redshift-dependence of the radial acceleration relation for particle dark matter
based on the numerical simulation of \cite{Keller}.  \label{fig3} }}
{
\includegraphics[width=8cm]{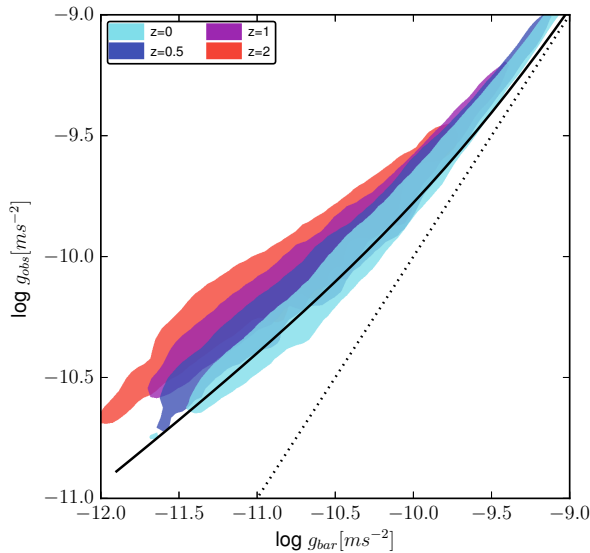}}
\end{figure}

This redshift-dependence induced through the changing
size of the cosmological horizon is, however, small compared to that expected from particle
dark matter. Figure \ref{fig2} shows the redshift-dependence 
of {\sc CEG} with Verlinde-matching and {\sc MOND}. This can be compared to
Figure \ref{fig3} (from \cite{Keller}) which was obtained
from the McMaster Unbiased Galaxy Simulations and predicts a much larger 
redshift-dependence than that of modified gravity.


Of course other numerical simulations might yield somewhat different results. Nevertheless we 
wish to propose here that, if data become better, the redshift-dependence of 
the radial acceleration could be used to tell apart modified gravity from particle dark matter.

\section*{Acknowledgements}

We thank Ben Keller for the permission to reuse Figure \ref{fig3} from \cite{Keller}.

\end{document}